\documentstyle[twocolumn,aps,prb,epsf]{revtex}
\begin{document}
\title{
The Spectrum-Generating Algebra of the Van Hove Scenario is SO(8)}

\author{R.S. Markiewicz$^{1,2}$ and M.T. Vaughn$^1$} 

\address{(1) Physics Department and (2) Barnett Institute, 
Northeastern U.,
Boston MA 02115}
\maketitle

\begin{abstract}
The various nesting and pairing instabilities of the generalized
Van Hove scenario can be classified via an $SO(8)$ spectrum-generating algebra.
An $SO(6)$ subgroup is an approximate symmetry group of the model, having two 
6-dimensional representations (`superspins').  This group contains as subgroups 
both the SO(5) and SO(4) groups found by Zhang, while one superspin is a 
combination of Zhang's 5-component superspin with a flux phase instability; the 
other includes a charge density wave instability plus s-wave superconductivity.
This is the smallest group which can describe both striped phases and 
superconductivity.
\end{abstract}


\narrowtext

Two groups play important roles in understanding a Hamiltonian: the {\it
symmetry group} allows a classification of its degenerate eigenstates, while the
Lie group of the {\it spectrum-generating algebra}\cite{SGA} (SGA) can be used 
to analyze the complete spectrum.  SGA's have proven to be useful in the study
of collective modes in nuclear and high-energy physics, while in condensed 
matter physics they have been used to study phase transitions in liquid He 
and in one-dimensional (1D) metals\cite{SoBir}.  In the 1D metals, the SGA is 
SU(8), with 63 elements and 56 possible order parameters including 
superconductivity and charge or spin density (CDW/SDW) waves.  This algebra has 
also been applied to the two-dimensional (2D) Hubbard model\cite{ErEiO}.  
However, we show that even for the generalized Hubbard model appropriate to the 
generalized Van Hove scenario, the appropriate SGA is SO(8), a considerably
smaller algebra.  This algebra has a natural subalgebra, SO(6), which acts as an
approximate symmetry group generalizing Zhang\cite{Zhang5}, and including both
his SO(5) and SO(4) as subgroups.  We identify SO(6) as the smallest group which
is capable of describing striped phases as well as superconductivity.
\par
For a one-dimensional (1D) metal\cite{Soly}, nesting involves the two points of 
the Fermi surface, at $\pm k_F$, with $k_F$ the Fermi momentum.  Since this
breaks momentum conservation (the states $+k_F$ and $-k_F$ become inequivalent),
the {\it full group} SU(8) must be taken as the SGA\cite{SoBir}.
On the other hand, in two dimensions, the dominant nesting arises at $\vec Q=
(\pi ,\pi )$, connecting the two Van Hove singularities (VHS's)\cite{RiSc} at
$(\pi ,0)$ and $(0,\pi )$.  Since the points $\pm(\pi ,0)$ are equivalent points
of the reciprocal space lattice, nesting singularities involve only order
parameters even in $\vec k$.  Hence the SGA is a proper subgroup of SU(8) --
the SO(8) algebra of Table I.  (Note that there is some ambiguity in defining a
SGA: here we define it as the algebra which contains the mean-field 
Hamiltonian.)
\par
\begin{figure}
\leavevmode
   \epsfxsize=0.38\textwidth\epsfbox{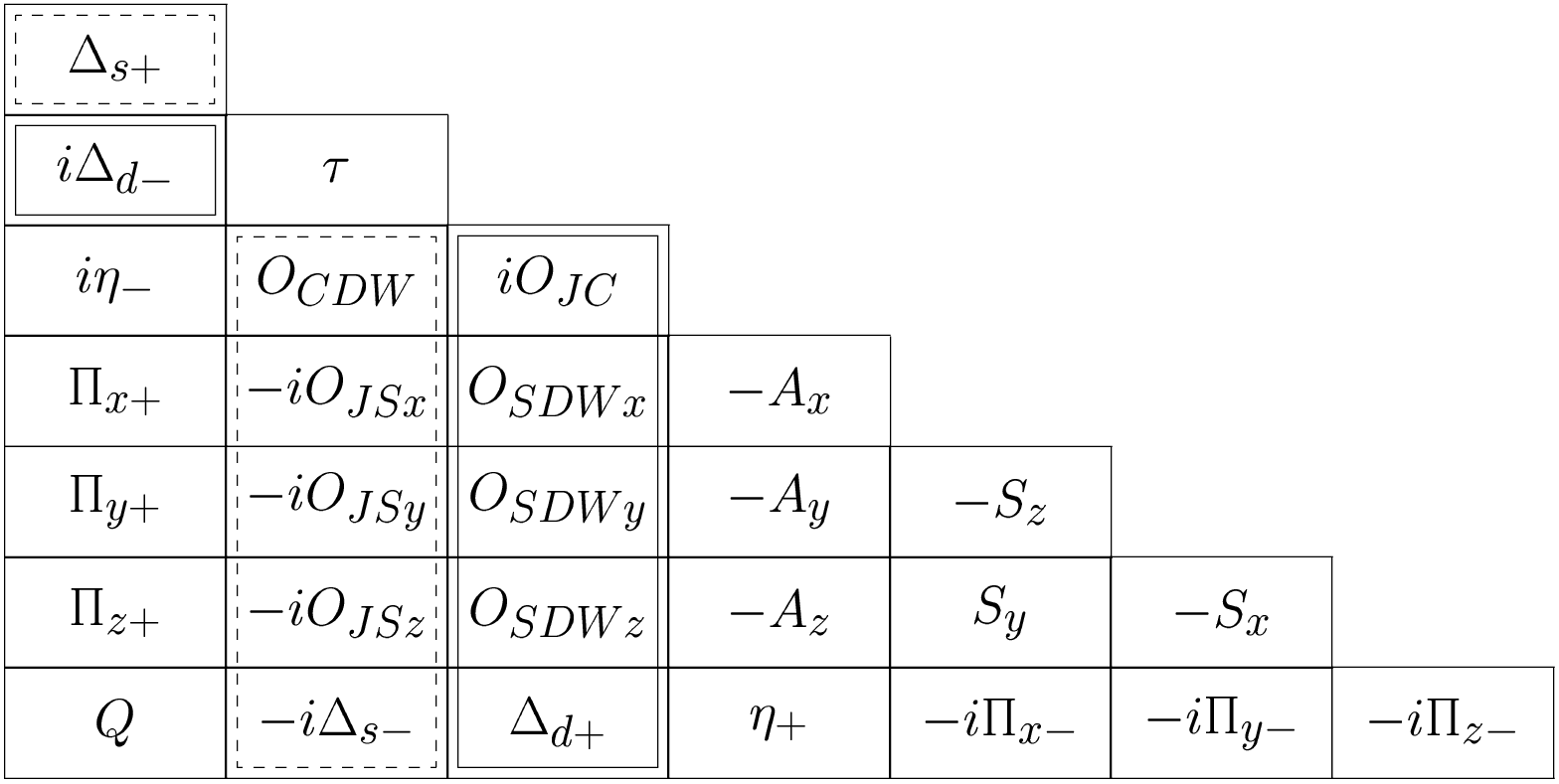}
\vskip-3.6cm 
\caption{Matrix Representation of SO(8), using the shorthand $O_{\pm}=
O\pm O^{\dagger}$.}
\label{fig:0}
\end{figure}
\par
There is a combinatoric interpretation of this SO(8) 
which is independent of any particular hamiltonian.
Consider an electronic system with a two-fold orbital degeneracy (labelled 1, 2)
in addition to the spin degeneracy.  The four creation operators can be written
as $C^{\dagger}_{1\uparrow}$, $C^{\dagger}_{2\uparrow}$, $C^{\dagger}_{1
\downarrow}$, and $C^{\dagger}_{2\downarrow}$.  Including both particle-hole
($C^{\dagger}C$) and particle-particle ($C^{\dagger}C^{\dagger}$ or $CC$)
operators, there are 28 pair operators, whose components define the Lie algebra 
of $SO(8)$ [recall that for $SO(N)$, the Lie algebra contains $N(N-1)/
2$ elements].  Particular linear combinations of these elements are listed in
Table I.  Figure~\ref{fig:0} rewrites these elements as an explicit 
representation of the Lie algebra of $SO(8)$.  The 28 generators are the 
antisymmetric matrices $L^{ij}$, with matrix elements $L^{ij}_{kl}=\delta^i_k
\delta^j_l-\delta^i_l\delta^j_k$.  Figure~\ref{fig:0} illustrates the 
equivalences as the lower half of an antisymmetric $L$-matrix.  The operators 
satisfy the Lie algebra, with standard $SO(8)$ commutation rules 
\begin{equation}
[L^{ij},L^{km}]=i(\delta_{ik}L^{jm}+\delta_{jm}L^{ik}-\delta_{im}L^{jk}-
\delta_{jk}L^{im}).
\label{eq:1}
\end{equation}
$SO(8-M)$ subalgebras can be formed by eliminating $M$ rows of the $L$ matrices,
along with their corresponding columns.  These will be designated as $\{I_1,...,
I_M\}$, where $I_1,...,I_M$ are the rows (and columns) which have been 
eliminated.  For instance, $\{234\}$ is the $SO(5)$ algebra studied by
Zhang\cite{Zhang5}.
\par
In a generalized Hubbard model\cite{Sch2}, the creation operators can be 
expanded in terms of operators localized near the corresponding VHS's:
\begin{equation}
a^{\dagger}_{i\sigma}\simeq{1\over 2}\bigl((-1)^{x_i}\psi^{\dagger}_{1\sigma}
(\vec r)+(-1)^{y_i}\psi^{\dagger}_{2\sigma}(\vec r)\bigr),
\label{eq:2}
\end{equation}
where $\psi^{\dagger}_{1\sigma}$ and $\psi^{\dagger}_{2\sigma}$ are slowly
varying functions of position $\vec r=a(x_i,y_i)$.  
A more precise definition is given  
\begin{tabular}{||c|c||}        %\hline
\multicolumn{2}{c}{{\bf Table I: Generators of SO(8) Lie Algebra}} \\ 
            \hline\hline
Operator & Representation \\   %\hline
    \hline\hline
$Q$ & $(C^{\dagger}_{1\uparrow}C_{1\uparrow}+C^{\dagger}_{2\uparrow}C_
{2\uparrow}+C^{\dagger}_{1\downarrow}C_{1\downarrow}+C^{\dagger}_{2\downarrow}C_
{2\downarrow})/2 - 1$   \\     \hline       
$\tau$ & $(C^{\dagger}_{1\uparrow}C_{1\uparrow}-C^{\dagger}_{2
\uparrow}C_{2\uparrow}+C^{\dagger}_{1\downarrow}C_{1\downarrow}-C^{\dagger}_
{2\downarrow}C_{2\downarrow})/2$   \\     \hline       
$S_z$ & $(C^{\dagger}_{1\uparrow}C_{1\uparrow}+C^{\dagger}_{2
\uparrow}C_{2\uparrow}-C^{\dagger}_{1\downarrow}C_{1\downarrow}-C^{\dagger}_{2
\downarrow}C_{2\downarrow})/2$   \\     \hline       
$A_z$ & $(C^{\dagger}_{1\uparrow}C_{1\uparrow}-C^{\dagger}_{2
\uparrow}C_{2\uparrow}-C^{\dagger}_{1\downarrow}C_{1\downarrow}+C^{\dagger}_{2
\downarrow}C_{2\downarrow})/2$   \\     \hline       
$S_x$ & $(C^{\dagger}_{1\uparrow}C_{1\downarrow}+C^{\dagger}_{2
\uparrow}C_{2\downarrow}+C^{\dagger}_{1\downarrow}C_{1\uparrow}+C^{\dagger}_{2
\downarrow}C_{2\uparrow})/2$   \\     \hline       
$A_x$ & $(C^{\dagger}_{1\uparrow}C_{1\downarrow}-C^{\dagger}_{2
\uparrow}C_{2\downarrow}+C^{\dagger}_{1\downarrow}C_{1\uparrow}-C^{\dagger}_{2
\downarrow}C_{2\uparrow})/2$   \\     \hline
$iS_y$ & $(C^{\dagger}_{1\uparrow}C_{1\downarrow}+C^{\dagger}_{2
\uparrow}C_{2\downarrow}-C^{\dagger}_{1\downarrow}C_{1\uparrow}-C^{\dagger}_{2
\downarrow}C_{2\uparrow})/2$   \\     \hline       
$iA_y$ & $(C^{\dagger}_{1\uparrow}C_{1\downarrow}-C^{\dagger}_{2
\uparrow}C_{2\downarrow}-C^{\dagger}_{1\downarrow}C_{1\uparrow}+C^{\dagger}_{2
\downarrow}C_{2\uparrow})/2$   \\     \hline
$O_{CDW}$ & $(C^{\dagger}_{1\uparrow}C_{2\uparrow}+C^{\dagger}_{2
\uparrow}C_{1\uparrow}+C^{\dagger}_{1\downarrow}C_{2\downarrow}+C^{\dagger}_{2
\downarrow}C_{1\downarrow})/2$   \\     \hline
$O_{SDWz}$ & $(C^{\dagger}_{1\uparrow}C_{2\uparrow}+C^{\dagger}_{2
\uparrow}C_{1\uparrow}-C^{\dagger}_{1\downarrow}C_{2\downarrow}-C^{\dagger}_{2
\downarrow}C_{1\downarrow})/2$   \\     \hline
$O_{JC}$ & $(C^{\dagger}_{1\uparrow}C_{2\uparrow}-C^{\dagger}_{2
\uparrow}C_{1\uparrow}+C^{\dagger}_{1\downarrow}C_{2\downarrow}-C^{\dagger}_{2
\downarrow}C_{1\downarrow})/2$   \\     \hline
$O_{JSz}$ & $(C^{\dagger}_{1\uparrow}C_{2\uparrow}-C^{\dagger}_{2
\uparrow}C_{1\uparrow}-C^{\dagger}_{1\downarrow}C_{2\downarrow}+C^{\dagger}_{2
\downarrow}C_{1\downarrow})/2$   \\     \hline
$O_{SDWx}$ & $(C^{\dagger}_{1\uparrow}C_{2\downarrow}+C^{\dagger}_{2
\uparrow}C_{1\downarrow}+C^{\dagger}_{1\downarrow}C_{2\uparrow}+C^{\dagger}_{2
\downarrow}C_{1\uparrow})/2$   \\     \hline       
$iO_{SDWy}$ &$(C^{\dagger}_{1\uparrow}C_{2\downarrow}+C^{\dagger}_{2
\uparrow}C_{1\downarrow}-C^{\dagger}_{1\downarrow}C_{2\uparrow}-C^{\dagger}_{2
\downarrow}C_{1\uparrow})/2$   \\     \hline
$O_{JSx}$ & $(C^{\dagger}_{1\uparrow}C_{2\downarrow}-C^{\dagger}_{2
\uparrow}C_{1\downarrow}+C^{\dagger}_{1\downarrow}C_{2\uparrow}-C^{\dagger}_{2
\downarrow}C_{1\uparrow})/2$   \\     \hline       
$iO_{JSy}$ & $(C^{\dagger}_{1\uparrow}C_{2\downarrow}-C^{\dagger}_{2
\uparrow}C_{1\downarrow}-C^{\dagger}_{1\downarrow}C_{2\uparrow}+C^{\dagger}_{2
\downarrow}C_{1\uparrow})/2$   \\     \hline       
\end{tabular}
\begin{tabular}{||c|c||c|c||}        %\hline
            \hline\hline
Op. & Representation &Op. & Representation \\   %\hline
    \hline\hline
$\Delta_s$ & $(C_{1\uparrow}C_{1\downarrow}+C_{2\uparrow}C_{2
\downarrow})/2$   &
$\Delta^{\dagger}_s$ & $(C^{\dagger}_{1\downarrow}C^{\dagger}_{1
\uparrow}+C^{\dagger}_{2\downarrow}C^{\dagger}_{2\uparrow})/2$   \\     \hline
$\Delta_d$ & $(C_{1\uparrow}C_{1\downarrow}-C_{2\uparrow}C_{2
\downarrow})/2$   &
$\Delta^{\dagger}_d$ & $(C^{\dagger}_{1\downarrow}C^{\dagger}_{1
\uparrow}-C^{\dagger}_{2\downarrow}C^{\dagger}_{2\uparrow})/2$   \\     \hline
$-i\Pi_y$ & $(C_{2\uparrow}C_{1\uparrow}+C_{2\downarrow}C_{1
\downarrow})/2$   &
$i\Pi^{\dagger}_y$ & $(C^{\dagger}_{1\uparrow}C^{\dagger}_{2\uparrow}+
C^{\dagger}_{1\downarrow}C^{\dagger}_{2\downarrow})/2$   \\     \hline
$\Pi_x$ & $(C_{2\uparrow}C_{1\uparrow}-C_{2\downarrow}C_{1
\downarrow})/2$   &
$\Pi^{\dagger}_x$ & $(C^{\dagger}_{1\uparrow}C^{\dagger}_{2\uparrow}-
C^{\dagger}_{1\downarrow}C^{\dagger}_{2\downarrow})/2$   \\     \hline
$\eta$ & $(C_{1\uparrow}C_{2\downarrow}+C_{2\uparrow}C_{1
\downarrow})/2$   &
$\eta^{\dagger}$ & $(C^{\dagger}_{2\downarrow}C^{\dagger}_{1\uparrow}+
C^{\dagger}_{1\downarrow}C^{\dagger}_{2\uparrow})/2$   \\     \hline
$\Pi_z$ & $(C_{1\uparrow}C_{2\downarrow}-C_{2\uparrow}C_{1
\downarrow})/2$   &
$\Pi^{\dagger}_z$ & $(C^{\dagger}_{2\downarrow}C^{\dagger}_{1\uparrow}
-C^{\dagger}_{1\downarrow}C^{\dagger}_{2\uparrow})/2$   \\     \hline
\end{tabular}
\vskip 0.1in
\begin{tabular}{||c||}        %\hline
\multicolumn{1}{c}{{\bf Table II: Interaction Terms}} \\ 
            \hline\hline
$(G_2-G_3)\Sigma_{\vec r}\Delta^{\dagger}_d(\vec r)\Delta_d(\vec r)$ \\  
   \hline       
$(G_2+G_3)\Sigma_{\vec r}\Delta^{\dagger}_s(\vec r)\Delta_s(\vec r)$ \\ 
    \hline       
$(2G_1+G_3-G_4)\Sigma_{\vec r}[O_{CDW}(\vec r)]^2$ \\     \hline       
$(G_3+G_4-2G_1)\Sigma_{\vec r}[O_{JC}(\vec r)]^2$ \\     \hline       
$(G_4-G_3)\Sigma_{\vec r}\vec O_{JS}(\vec r)\cdot\vec O_{JS}(\vec r)$ \\  
   \hline       
$-(G_3+G_4)\Sigma_{\vec r}\vec O_{SDW}(\vec r)\cdot\vec O_{SDW}(\vec r)$ \\ 
    \hline       
\end{tabular}
\vskip 0.1in
\par\noindent
in Refs.~\cite{RM8c,MarV}.  The Lie algebra of Table I corresponds to $O
\rightarrow\Sigma_{\vec r}O(\vec r)$, with $C^{\dagger}_{i\sigma}\rightarrow
\psi^{\dagger}_{i\sigma}(\vec r)$, etc.  With this definition, the operators 
become equivalent to those introduced by Schulz\cite{Sch2} and 
Zhang\cite{Zhang5}.  The SGA {\bf G} is defined in Fourier space as ${\bf G}=
\oplus_{\vec k}{\bf g}_{\vec k}$, where ${\bf g}_{\vec k}$ is the algebra of a 
particular $\vec k$-component of the Fourier transformed operators of Table I.
\par
The interaction terms in the 
generalized Hubbard hamiltonian\cite{Sch2} can be written in terms of pairs
of these operators, Table II.  Here the $G_i$'s are coupling 
constants, which can be related to the Hubbard $U$ and to various near-neighbor 
interaction terms\cite{Sch2}.  For the pure Hubbard model, $G_1=G
_2=G_3=G_4=U/4\pi t$ and $t$ is the nearest neighbor hopping parameter.  The 
form of the interaction term is not unique, since a number of alternative terms 
can arise by anticommuting the operators.
\begin{figure}
\leavevmode
   \epsfxsize=0.5\textwidth\epsfbox{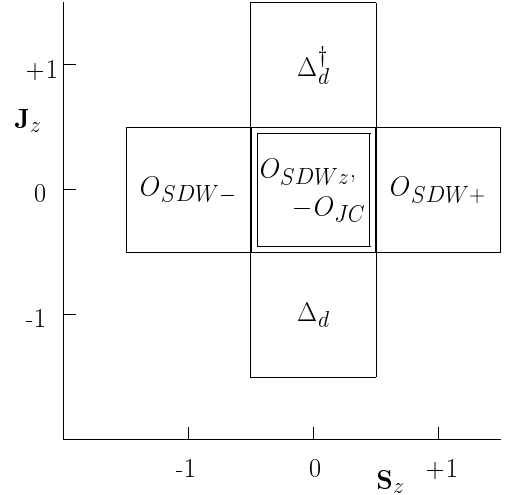}
\vskip-3.6cm 
\caption{$SO(4)$ weight diagram of {\bf V$^{\prime}$}, where $S_z$ ($J_z$) is 
the z component of spin (pseudospin).}
\label{fig:1}
\end{figure}
\par
While SO(8) is the SGA of the generalized Hubbard model, Zhang's SO(5) is an
(approximate) symmetry algebra of the same model -- in particular, the
collective modes (SDW and d-wave superconductivity for SO(5)) are {\it
elements} of the SGA, but are {\it components} of a superspin, which transforms
under the symmetry group.  For the generalized Hubbard model, the natural 
(approximate) symmetry group is SO(6), defined as follows.
$SO(8)$ contains an $SO(3)$ algebra, which we call {\it isospin}, generated by 
$T_0=\tau$, 
$T_{\pm}=O_{CDW}\pm O_{JC}$.  This algebra is the algebra of the VHS's: the 
z-component of the isospin, $T_0$, measures the excess population of the 1-VHS 
over the 2-VHS.  Those operators which do not commute with $T_0$ can lead to a 
nesting or pairing instability.
Hence, the important transformation group of the VHS's is the $SO(6)$ subgroup 
$\{23\}$ which commutes with $T_0$, leading to the decomposition scheme: 
\begin{equation}
{\bf SO(8)}\rightarrow{\bf SO(6)}_{\{23\}}\oplus\bf{V_+}
\oplus\bf{V_-}\oplus\bf{\tau},
\label{eq:13}
\end{equation}
under which ${\bf{28}}\rightarrow ({\bf{15}},0)+({\bf{6}},1)+({\bf6},-1)
+({\bf1},0)$, where $({\bf m},n)$ denotes representation {\bf m} of $SO(6)$ and 
eigenvalue $n$ of $T_0$.  The 6-vectors can be denoted ${\bf V_{\pm}}={\bf V}
\pm{\bf V^{\prime}}$ with
\begin{eqnarray}
{\bf V} =\{L_{21},L_{42},L_{52},L_{62},L_{72},L_{82}\}\nonumber \\
{\bf V^{\prime}}=\{L_{31},L_{43},L_{53},L_{63},L_{73},L_{83}\},
\label{eq:15}
\end{eqnarray}
shown boxed in Fig.~\ref{fig:0}.  The group structure of {\bf V$^{
\prime}$} is shown in Fig.~\ref{fig:1}, where $J_z$ is the 
z-component of the pseudospin operator introduced by Yang and Zhang\cite{Zhang4}
and $O_{SDW\pm}=\mp(O_{SDWx}\pm iO_{SDWy})/\sqrt{2}$.
An analogous diagram can be drawn for {\bf V}.
The group $SO(6)_{\{23\}}$ transforms the components of each of these 6-vectors 
among themselves, without mixing the two vectors, while $\tau$ transforms the
vectors into each other.
\par
A number of points should be noted. 
(1) The $SO(6)$ group $\{23\}$ contains Zhang's $SO(5)$ group as a subgroup, as
well as the the $SO(4)$ group introduced by Yang and Zhang\cite{Zhang4}.
Moreover, {\bf V$^{\prime}$}, Fig. \ref{fig:1}, combines Zhang's $SO(5)$ 
superspin with $O_{JC}$, which is essentially equivalent to the flux
phase\cite{Affl}.  
\par
(2) The twelve components of superspin are precisely the collective modes 
identified earlier by Schulz\cite{Sch2}, and most of them have been found to
play an important role in the cuprates: s-wave superconductivity in 
electron-doped and (possibly) overdoped cuprates, CDW's near optimal 
doping\cite{Surv,Pstr}, the flux phase near half filling\cite{flux,Pstr}.
\par
(3) For a bare band dispersion (neglecting interactions) of the form 
\begin{equation}
\epsilon_{\vec k}=-2t(\cos k_xa+\cos k_ya)+4t^{\prime}\cos k_xa\cos k_ya,
\label{eq:24}
\end{equation}
two parameters control the symmetry of the quadratic part of the generalized 
Hubbard hamiltonian, $t^{\prime}$ and $\tilde\mu =E_F-E_V$, the shift of the
Fermi level $E_F$ from the VHS $E_V$.  When both parameters are zero (half
filling with square Fermi surface) the hamiltonian has an extra pseudospin 
symmetry\cite{Zhang4}.  In this case, the nature of the ground state instability
is controlled solely by the interaction terms (the $G$'s).  A pure Hubbard 
interaction ($U$) breaks the $SO(6)$ symmetry (Table II):
\begin{equation}
\bf SO(6)\rightarrow SO(3)\oplus SO(3), 
\label{eq:14}
\end{equation}
with one $SO(3)$ ordinary spin, and the other the pseudospin\cite{Zhang4}.
Both 6-vectors are broken down to pairs of 3-vectors 
\begin{eqnarray}
\bf V\rightarrow \{O_{JSi}\}\oplus \{\Delta^{\dagger}_s, \Delta_s, 
O_{CDW}\}\nonumber \\
\bf{V^{\prime}}\rightarrow \{O_{SDWi}\}\oplus
\{\Delta^{\dagger}_d, \Delta_d, O_{JC}\};
\label{eq:16}
\end{eqnarray}
however, there remains an accidental degeneracy of one vector ($O_{JS}$) 
with the opposite pseudovector.  At half filling the lowest energy state is 
$O_{SDW}$.  As discussed below, this weak coupling result must be corrected for
strong correlation effects.
\par
When the Fermi surface is distorted from square, either by doping away from
half filling ($\tilde\mu$) or by introducing second-neighbor hopping terms 
$t^{\prime}$, the pseudospin degeneracy is broken, in such a way as to {\it 
favor pairing over nesting instabilities}.  This can be seen by 
Hartree-Fock\cite{BFal} or renormalization group\cite{Dzy,Sch,Surv} analyses 
or by a linear response analysis (following Ref.~\cite{Mor}).
\par
If the superspin is written as $\vec O$ (a twelve component vector incorporating
both representations), then in linear response theory it is assumed 
that there is an applied field $\vec h_O$ (also a 12-vector) which couples to 
$\vec O$.  The hamiltonian in the presence of $\vec h_O$ is
\begin{equation}
H=\Sigma_{\vec k\sigma}\epsilon_{\vec k}a^{\dagger}_{\vec k\sigma}a_{\vec k
\sigma}+\vec h_O\cdot\vec O
\label{eq:17}
\end{equation}
(here the terms in $G$ have been neglected), with resulting free energy 
\begin{eqnarray}
F_0(\vec O)=\Omega_0(\mu ,\vec O,T)+\mu N \nonumber \\
\Omega_0(\mu ,\vec O,T)=-2k_BT\Sigma_{\vec k\sigma}\ln (1+e^{-(E_{\vec k\sigma}-
\mu )/k_BT}) \nonumber \\
+\vec h_O\cdot\vec O,
\label{eq:18}
\end{eqnarray}
with $E_{\vec k\sigma}$ the quasiparticle energy found by applying a
Bogoliubov-Valentin transformation to Eq.~\ref{eq:17} or via SGA
techniques\cite{SGA,SoBir}.  The expectation value of each superspin component 
$O_i$ is found from
\begin{equation}
{\partial\Omega_0\over\partial h_{Oi}}=0,
\label{eq:19}
\end{equation}
and the corresponding susceptibility is
\begin{equation}
\chi_{0i}=\lim_{\vec h_O\rightarrow 0}({O_i\over h_{Oi}}).
\label{eq:20}
\end{equation}
Including the interaction terms, the Hartree-Fock free energy becomes
\begin{equation}
F_{HF}(\vec O)=\Sigma_i\Bigl({1\over 2\chi_{0i}}+G_i\Bigr)O_i^2,
\label{eq:21}
\end{equation}
leading to an instability of the $i$th mode when 
\begin{equation}
1+2\chi_{0i}G_i=0.
\label{eq:22}
\end{equation}
If the quadratic hamiltonian is symmetric under $SO(6)$, then the component with
the most negative $G_i$ is the first to diverge.  For finite $\tilde\mu$ or
$t^{\prime}$,the hamiltonian still preserves particle number, so there are only
two independent susceptibilities, the particle-hole
susceptibility $\chi_{00}$ and the pair susceptibility $\chi_{02}$, with
\begin{eqnarray}
\chi_{00}=-2\Sigma_{\vec k\sigma}{f(\epsilon_{\vec k\sigma})\over\epsilon_{\vec 
k\sigma}-\epsilon_{\vec k+\vec Q,\sigma}} \nonumber \\
\chi_{02}=-\Sigma_{\vec k\sigma}{f(\epsilon_{\vec k\sigma})\over\epsilon_{\vec k
\sigma}-\epsilon_F}.
\label{eq:23}
\end{eqnarray}
Note that in nearest-neighbor hopping models ($t^{\prime}=0$) $\epsilon_{\vec k+
\vec Q,\sigma}=-\epsilon_{\vec k\sigma}$, and the two expressions become 
equivalent when $\epsilon_F=0$ -- i.e., at half filling.
\par
\begin{figure}
\leavevmode
   \epsfxsize=0.33\textwidth\epsfbox{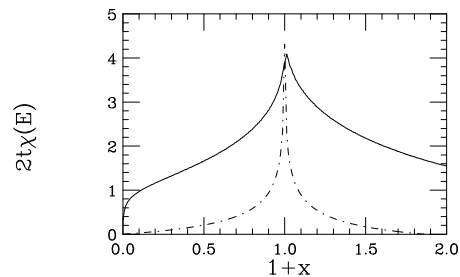}
\vskip0.5cm 
\caption{Susceptibilities $\chi_{00}$ (dotdashed line) and $\chi_{02}$ (solid
line) vs. band filling $1+x$ for Eq.~\protect\ref{eq:24} with $t^{\prime}=0$.}
\label{fig:2}
\end{figure}
\begin{figure}
\leavevmode
   \epsfxsize=0.33\textwidth\epsfbox{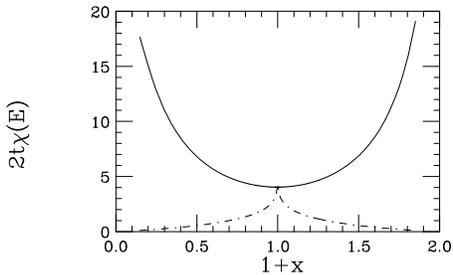}
\vskip0.5cm 
\caption{As in Fig.~\protect\ref{fig:2}, but with $4t^{\prime}=E_F$.}
\label{fig:4}
\end{figure}

Figure~\ref{fig:2} illustrates the doping dependence of these susceptibilities 
for a Hubbard band with nearest-neighbor hopping only ($t^{\prime} =0$).
The point of maximum instability (largest $\chi$) coincides with the point at 
which the VHS 
\par\noindent
crosses the Fermi level -- half filling when $t^{\prime}=0$.
When $x=t^{\prime}=0$, the susceptibilities are degenerate, $\chi_{00}=
\chi_{02}$, as expected from the pseudospin symmetry\cite{Zhang4}.
However, as soon as the system is doped away from half filling
($x\ne 0$) the electron-hole susceptibility drops precipitously, whereas the
pair susceptibility falls off much more gradually.  A similar effect arises if
the system is maintained at optimal doping (the VHS), but the parameter $t^
{\prime}$ is varied -- indeed $\chi_{02}$ actually increases with increasing 
$t^{\prime}$, Fig~\ref{fig:4}.  This striking difference is readily understood: 
the electron-hole susceptibility involves inter-VHS nesting, which gets
progressively worse as the Fermi surface gets more curved, whereas the
electron-electron susceptibility involves intra-VHS scattering, and increases
with $t^{\prime}$ as the Fermi surfaces become nearly 1D near the VHS's.  (In 
Figs.~\ref{fig:2}-\ref{fig:4}, the logarithmic divergence at the VHS was cut off
by adding a small imaginary term to the denominator of $\chi$.)
\par
Figure~\ref{fig:2} is consistent with the RG results of Schulz\cite{Sch2}. From
Eq.~\ref{eq:22}, when all $\chi$'s are equal the order parameter associated 
with the most negative $G$ is the first to go singular.  For the pure Hubbard
model, this means that the leading instability at half filling is the SDW.
When the material is doped, the SDW susceptibility plummets, and at some point
d-wave superconductivity becomes favorable.  In agreement with 
Zhang\cite{Zhang5}, the shift of the Fermi energy from the VHS is a relevant 
parameter in driving this SDW$\rightarrow$ d-wave superconducting transition.
\par
Despite the simplicity of this picture, a purely SO(5) model cannot explain the 
full physics of the cuprates.  First, the above analysis is in the weak coupling
limit, and a strong coupling reanalysis of Table II (J-term dominant) shows that
the flux phase -- not included in SO(5) -- is the lowest energy state.
A second problem is in La$_{2-x}$Ba$_x$CuO$_4$ and La$
_{2-x-y}$Nd$_y$Sr$_x$CuO$_4$, near $x=1/8$, where the striped phase is 
commensurately pinned, leading to long-ranged magnetic and charge 
order\cite{Tran}.  At the same time, superconductivity is strongly suppressed, 
demonstrating that whatever the driving force for charge order may be, it is not
superconductivity, but is in competition with superconductivity.  Since SO(5)
only allows for antiferromagnetism and superconductivity, it does not have
sufficient flexibility to properly describe this situation.  There are strong
hints that the charged stripes are associated with a CDW: the low-temperature
tetragonal phase is nearly coterminus with the long-range SDW-ordered phase,
and the fact that the charged stripes are best seen by neutron diffraction
suggests a strong associated lattice distortion.
There is considerable additional evidence that
phonons and structural instabilities play an important role in
the doped material\cite{Surv}.  Hence, for a detailed description of the doping
dependence of the pseudogap, striped phases, and extended VHS's, it may be 
necessary to recognize that strong electron-phonon coupling can lead to a 
crossover to a groundstate involving the {\bf V} 6-vector\cite{Pstr}.  
\par
MTV's research is supported by the Dept. of Energy under Grant \#
DE-FG02-85ER40233.  Publication 722 of the Barnett Institute.

\end{document}